\documentclass[aps,pra,twocolumn,superscriptaddress,showpacs]{revtex4}

\usepackage{graphicx}
\usepackage{epsfig}
\usepackage{subfigure}

\usepackage{amsmath,amsthm,amsfonts,latexsym}

\newcommand{\ket}[1]{| #1 \rangle}

\newcommand{\mtx}[2]{\left(\begin{array}{#1}#2\end{array}\right)}

\begin{document}


\title{Geometric local invariants and pure three-qubit states}


\author{Mark S. Williamson}\email{m.s.williamson04@gmail.com}
\affiliation{Centre for Quantum Technologies, National University of Singapore, 3 Science Drive 2, Singapore 117543,}
\affiliation{Erwin Schr\"{o}dinger International Institute for Mathematical Physics, Boltzmanngasse 9, 1090 Wien, Austria,}
\author{Marie Ericsson}
\affiliation{Department of Quantum Chemistry, Uppsala University, Box 518, SE-751 20 Uppsala, Sweden,}
\author{Markus Johansson}
\affiliation{Department of Quantum Chemistry, Uppsala University, Box 518, SE-751 20 Uppsala, Sweden,}
\author{Erik Sj\"{o}qvist}
\affiliation{Centre for Quantum Technologies, National University of Singapore, 3 Science Drive 2, Singapore 117543,}
\affiliation{Department of Quantum Chemistry, Uppsala University, Box 518, SE-751 20 Uppsala, Sweden,}
\author{Anthony Sudbery}
\affiliation{Department of Mathematics, University of York, Heslington, York YO10 5DD, UK,}
\author{Vlatko Vedral}
\affiliation{Centre for Quantum Technologies, National University of Singapore, 3 Science Drive 2, Singapore 117543,}
\affiliation{Clarendon Laboratory, University of Oxford, Parks Road, Oxford OX1 3PU, UK,}
\author{William K. Wootters}
\affiliation{Department of Physics, Williams College, Williamstown, Massachusetts 01267, USA.}

\date{\today}

\begin{abstract}
We explore a geometric approach to generating local
$SU(2)$ and $SL(2,\mathbb{C})$ invariants for a collection of qubits inspired by lattice
gauge theory. Each local invariant or `gauge' invariant is
associated to a distinct closed path (or plaquette) joining some or
all of the qubits. In lattice gauge theory, the lattice points are
the discrete space-time points, the transformations between the
points of the lattice are defined by parallel transporters and the
gauge invariant observable associated to a particular closed path is
given by the Wilson loop. In our approach the points of the lattice
are qubits, the link-transformations between the qubits are defined
by the correlations between them and the gauge invariant
observable, the local invariants associated to a particular
closed path are also given by a Wilson loop-like construction. The link
transformations share many of the properties of parallel
transporters although they are not undone when one retraces one's
steps through the lattice. This feature is used to generate many of
the invariants. We consider a pure three
qubit state as a test case and find we can generate a
complete set of algebraically independent local invariants in this
way, however the framework given here is applicable to generating local unitary invariants for mixed states composed of any number of $d$ level quantum systems. We give an operational interpretation of these invariants in terms of observables.
\end{abstract}


\maketitle

\section{Introduction}

One approach to the study of entanglement is the identification of
local invariants of a collection of quantum objects. With this
approach we imagine the distant labs scenario in which $N$ spatially
separated parties each hold one of the subsystems of an $N$ particle
entangled state in their laboratory and they are free to make
arbitrary transformations on their subsystem. One then looks
for properties of the state that remain unchanged under such
local transformations since, under the conditions that the transformations are unitary, entanglement is
defined to be invariant. If the transformations belong to the group $SL(2,\mathbb{C})$ it turns out that entanglement, given by the well known measure concurrence, is also invariant. Rephrasing, we can write this scenario as a
non-abelian lattice gauge theory; the arbitrary transformations are
non-abelian local gauge transformations made on $N$ subsystems, the
$N$ points of the lattice. Entanglement is a gauge
invariant observable of the theory. It is this similarity that
inspires our work.

Quite a lot is known about the local unitary invariants of simple entangled
states. For example, for a pure state of a pair of qubits, there is
essentially only one local invariant (not counting the
normalization); it characterizes the amount of entanglement between
the two qubits. For a pure state of three qubits, one can identify
five independent local invariants, four of them expressing a
different aspect of the state's entanglement \cite{ref:Carteret00,ref:Linden&Popescu98,ref:Sudbery01}. A fifth, the Kempe
invariant, is not well understood \cite{ref:Kempe99}. There exist well known algebraic
methods for generating invariants
\cite{ref:Grassl98,ref:Rains00,ref:Barnum01,ref:Teodorescu03,ref:Osterloh&Siewert05,ref:Osterloh08},
but as the number of subsystems increases, the problem of
identifying and interpreting independent invariants rapidly becomes
very complicated.

Here we explore a different approach inspired by lattice gauge theory \cite{ref:Muenster&Walzl00}. For a collection of $N$ qubits, we
consider any closed path connecting some of the qubits, and we
associate an invariant quantity with each such path. The invariant
is formed by taking the trace of a transformation associated with
the closed path, which in turn is the product of transformations
associated with the individual two qubit links. Each of these
`link-transformations' is determined by the density matrix of the
two qubits connected by the link. Because this density matrix will
typically change if one performs a local operation on either
of the two qubits, each link-transformation will also typically
change under such local operations.  The overall transformation
around a closed loop can similarly change as one performs a local operation on the qubit
that defines the loop's starting point. However, the \emph{trace} and the eigenvalues of
the overall transformation do not change under any single-qubit
operations. The trace is our invariant. In fact we will generate a
few distinct invariants associated with the same path, by using
different, but closely related, ways of making the correspondence
between a two-qubit density matrix and a link-transformation i.e. one has the choice whether to apply a spin flip to each qubit in a given loop.

Other authors have explored relations between entanglement and gauge
transformations, in the context of an analysis of the geometry of
the set of states
\cite{ref:Mosseri01,ref:Bernavig03,ref:Levay04,ref:Levay05}. Our
approach is different in that the paths we consider are not paths in
the set of states but discrete paths connecting the qubits
themselves.

Thus our invariants are determined once we specify the
correspondence between a two qubit density matrix and a
link-transformation.  The first rule we consider, and the one from
which the other cases are derived, is this one:
\begin{equation}\label{sec:SU2invariants:eq:intro}
M_b = \text{tr}_a\left[ (M_a \otimes \mathbf{I}_b) \rho_{ab}\right].
\end{equation}
Here $\rho_{ab}$ is the two qubit density matrix in question, $M_a$
is a $2\times 2$ Hermitian matrix, and $M_b$ is its image under our
transformation.  (In section~\ref{sec:SU2invariants:path} we
interpret this rule in terms of local observables.)

We hope that this geometric approach will ultimately prove useful in
generating and classifying invariants of systems with many parts. In
this paper we try out our ideas by applying them to a simple
system of three qubits in a pure state.  For that case, as indicated
above, a natural set of $SU(2)$ invariants is already known.  We ask
whether this set, or an equivalent set, can be generated via our
construction. We also ask whether the path-based approach sheds any
light on the physical meaning of these invariants.

In the following section we introduce our basic path-based method of
generating invariants.
Section~\ref{sec:SU2invariants:identification} applies this idea to
pure states of three qubits and makes the connection between the
invariants generated by this method and the standard three qubit
$SU(2)$ invariants that have been identified previously. In section~\ref{sec:SU2invariants:SL2} we show how to generate $SL(2,\mathbb{C})$ invariants by simply spin flipping each qubit in a loop. In section~\ref{sec:SU2invariants:interpretation} we give an operational interpretation of these invariants in terms of observables. Finally, we draw
conclusions in section~\ref{sec:SU2invariants:conclusion} and outline how one would extend this approach to mixed states comprised of any number of qu$d$its.

\section{Path-based invariants}\label{sec:SU2invariants:path}

Our basic method of associating a transformation with each two qubit
link is motivated by a thought experiment.  Imagine many $N$ qubit
systems, each having distinguishable qubits labeled $a$, $b$, $c$,
$\ldots$, and each system being in the same quantum state
$\rho$.  We use $\rho$ to define a transformation
from qubit $a$ to qubit $b$ as follows.  On several copies of the state $\rho$, perform a general quantum measurement on qubit $a$, one of whose
outcomes is represented by the operator $M_a$.  (This operator is
arbitrary except that it must be positive semi-definite and less than the identity
so that it can be part of a legitimate measurement.)  Now consider
only those instances of qubit $a$ for which this particular outcome
is actually achieved.  In those cases, the state of qubit $b$ has
been `collapsed' into some state, typically mixed, even though qubit
$b$ has not interacted with the measuring device.  The final state
of qubit $b$ is in fact proportional to
\begin{equation}\label{sec:SU2invariants:eq:testing}
M_b = \text{tr}_a [(M_a \otimes \mathbf{I}_b)\rho_{ab}],
\end{equation}
where $\rho_{ab}$ is the original reduced density matrix of qubits
$a$ and $b$ when the whole system is in state $\rho$.  The
normalization of $M_b$, that is, $\text{tr}M_b$, is equal to the
\emph{probability} of getting the outcome represented by $M_a$.  In
this way the density matrix $\rho_{ab}$ defines a linear
transformation from operators on qubit $a$ to operators on qubit
$b$, namely, the transformation that takes $M_a$ to $M_b$.  It is
convenient to represent this linear transformation as a matrix by
writing $M_a$ and $M_b$ in terms of Pauli spin matrices. Let the
four real numbers $m^{a}_i$, $i=0,\ldots, 3$, be defined by
\begin{equation}
M_a = m^{a}_0 \sigma_0 + m^{a}_1 \sigma_1 + m^{a}_2 \sigma_2 +
m^{a}_3 \sigma_3,
\end{equation}
where $\sigma_0$ is the $2\times 2$ identity matrix and the other
$\sigma_i$'s are the usual Pauli matrices, and let the components of
$M_{b}$ be defined similarly. Then we can express our transformation
as the $4\times 4$ matrix $S(b,a)$ such that
\begin{equation}
\mathbf{m}^{b} = S(b,a) \mathbf{m}^{a},
\end{equation}
where $\mathbf{m}^{a}$ and $\mathbf{m}^{b}$ are column 4-vectors with
components $m^{a}_i$ and $m^{b}_i$. Writing out
eq.~(\ref{sec:SU2invariants:eq:testing}) explicitly in this operator
basis, we have
\begin{equation}
\sum_k m^{b}_k\sigma^{b}_k = \text{tr}_a \left[\left(\sum_i m^{a}_i
\sigma^{a}_i \otimes \mathbf{I}_b \right) \rho_{ab}\right].
\end{equation}
Multiplying both sides by $\sigma_j^{b}$ and tracing over qubit $b$,
we get an explicit expression for the components of the matrix
$S(b,a)_{ji}$
\begin{equation}\label{sec:SU2invariants:eq:Sdef}
S(b,a)_{ji} = \frac{1}{2} \text{tr}\left( \sigma^{a}_i\otimes
\sigma^{b}_j \rho_{ab}\right) = \frac{1}{2}\langle
\sigma^{a}_i\otimes \sigma^{b}_j \rangle.
\end{equation}
So in this representation, the matrix representing our
transformation from qubit $a$ to qubit $b$ is proportional to the
spin correlation matrix. Our link-transformations are specified by
the correlations between the qubits joined by the link.

We can now imagine repeating this process at qubit $b$.  That is,
starting with several pristine, unmeasured copies of the state
$\rho$, we imagine performing a measurement on qubit $b$,
one of whose outcomes is represented by the same operator $M_b$ that
was the result of the first measurement.  When this outcome is
achieved, qubit $c$ will be collapsed into some mixed state
proportional to $M_c$ defined as in
eq.~(\ref{sec:SU2invariants:eq:testing}).

\begin{figure}
\begin{center}
\includegraphics[width=7cm]{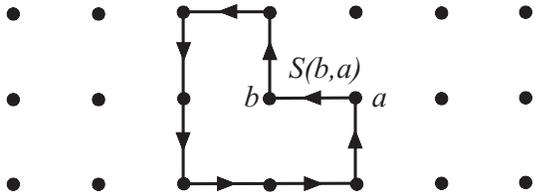}
\end{center}
\caption[Local invariants as distinct paths in on a lattice.]{A
schematic of our method of generating local invariants or `gauge
invariants' for a collection of quantum objects. The idea is
inspired by lattice gauge theory with the lattice points
representing qubits and the transformations $S(b,a)$ between lattice
points $a$ and $b$ given by the correlations between qubits $a$ and
$b$. Each distinct closed path or loop generates a local
invariant.}\label{sec:SU2invariants:fig:schematic}
\end{figure}

Continuing in this way around a closed loop, we finally collapse
qubit $a$ into some state proportional to
\begin{equation}
M'_a = \text{tr}_z\,[(M_z \otimes \mathbf{I}_a)\rho_{za}],
\end{equation}
where qubit $z$ is the one that precedes qubit $a$ at the end of the
loop.  In this way we have mapped, via the whole loop $\mathcal{C}$,
an operator $M_a$ on the state space of qubit $a$ into another
operator $M'_a$ on the same space.  The matrix representing this
transformation is
\begin{equation}\label{sec:SU2invariants:eq:totaltrans}
S(a,a;\mathcal{C}) = S(a,z)\cdots S(c,b)S(b,a).
\end{equation}
In other words the overall transformation taking our initial
measurement four vector $\mathbf{m}^a$ around the closed loop back
to our new four vector $\mathbf{m}^{a'}$ is
\begin{equation}
\mathbf{m}^{a'}=S(a,a;\mathcal{C})\mathbf{m}^a.
\end{equation}
As we show in the following section, the trace of this matrix is
invariant under all single qubit unitary transformations. This trace
is the invariant we associate with the given closed path. We present
a graphical illustration of the idea in
figure~\ref{sec:SU2invariants:fig:schematic}.

Our basic transformation, eq.~(\ref{sec:SU2invariants:eq:testing}),
is reminiscent of the transformation that would be associated with
the two qubit state $\rho_{ab}$ by the Jamio\l kowski isomorphism
\cite{ref:Jamiolkowski72}, which provides a general correspondence
between bipartite states and trace-preserving operations. The
transformation defined by that isomorphism would be
\begin{equation}\label{sec:SL2invariants:eq:Jam}
M^{J}_b = 2 \text{tr}_a [(M^T_a \otimes \mathbf{I}_b)\rho_{ab}] .
\end{equation}
That is, it would be normalized differently and it would require
taking the transpose of the initial operator.  The transpose is
included in order to make the transformation a legitimate quantum
operation---specifically, in order to make it completely positive.
In contrast, the transformation defined by our
eq.~(\ref{sec:SU2invariants:eq:testing}) need not be completely
positive. We have chosen the form of
eq.~(\ref{sec:SU2invariants:eq:testing}) as we have because our aim
is not to define a quantum operation but rather to generate
invariants. If we had included the transpose, the resulting
$\text{tr} S(\mathcal{C})$ would not have been invariant. Moreover,
even though our transformation is not a quantum operation, it does
have a physical interpretation in terms of measurement.

Though introducing the transpose would spoil the invariance, there
is a closely related operation that does not have this effect,
namely, the spin flip. At any point along a
closed path, we have the option of inserting a spin flip without
destroying the invariance. In our thought experiment, this would
mean that, having collapsed, say, qubit $b$ into the (subnormalized)
state $M_b$, in our next step we would perform a measurement with an
outcome represented by $\tilde{M}_b$, where the tilde represents the
spin flip.  That is,
\begin{equation}
\tilde{M} = \sigma_2 M^T \sigma_2.
\end{equation}
The effect of a spin flip on the vector $m$ representing $M$ in the
Pauli basis is simply to multiply $m_1$, $m_2$, and $m_3$ by $-1$
and to leave $m_0$ unchanged. ($\sigma_2$ multiplies $m_1$ and $m_3$ by $-1$ while the transposition, equivalent to complex conjugation, multiplies $m_2$ by $-1$. The spin flip is anti-unitary and not a physical operation). That is, in this representation a
spin flip is represented by the matrix $\eta$, the Minkowski metric;
\begin{equation}
\eta = \mtx{cccc}{1 & 0 & 0 & 0 \\ 0 & -1 & 0 & 0 \\ 0 & 0 & -1 & 0 \\
0 & 0 & 0 & -1}.
\end{equation}
We will label each of our invariants by the closed path that defines
it, indicating with a tilde any site at which we have added a spin
flip.  Thus, for example, $I(a\tilde{b}c)$ is the invariant defined
by
\begin{equation}
I(a\tilde{b}c) = \text{tr}\left\{ S(a,c)S(c,b)\eta S(b,a)\right\}.
\end{equation}

It is not hard to see that a spin flip indeed preserves the $SU(2)$ invariance. In the following section, we will show
local transformations $\mathcal{U}$ become elements of $SO(3)$
acting only on the spatial dimensions of $S$, those with index
values $1,2,3$ and not on the dimension associated with the
identity. That is, they are block-diagonal matrices with a $1\times
1$ block and a $3\times 3$ block.  Thus they commute with $\eta$ and
therefore still cancel each other.

In fact we find that the inclusion of a spin flip on every qubit in a path results in not only an $SU(2)$ invariant but also a $SL(2,\mathbb{C})$ invariant, a group which contains $SU(2)$. The group $SL(2,\mathbb{C})$ represents the most general, local operations, such as Kraus operations, that one may perform on a qubit up to a positive constant less than unity. This stronger invariance is interesting as the well known entanglement measures concurrence and three-tangle exhibit this higher invariance \cite{ref:Verstraete01a}. We demonstrate this in the following section.

\section{Properties of
link-transformations}\label{sec:SU2invariants:properties}

\subsection{Local operations}

Suppose that on one qubit, say qubit $b$ for definiteness, we
perform a general local operation, not necessarily unitary
$\mathcal{U}_b$ i.e.
\begin{equation}
M_b = \text{tr}_a [(M_a \otimes \mathbf{I}_b)(\mathbf{I}_a \otimes
\mathcal{U}_b)\rho_{ab} (\mathbf{I}_a \otimes \mathcal{U}_b^\dag)].
\end{equation}
In a cycle that includes the links $a\rightarrow b$ and
$b\rightarrow c$, this transformation would change both $S(b,a)$ and
$S(c,b)$.  For example, $S(b,a)_{ji}$ would be transformed into
\begin{equation}
\frac{1}{2}\langle \sigma^{a}_i\otimes ({\mathcal{U}_b}^\dag
\sigma^{b}_j \mathcal{U}_b )\rangle.
\end{equation}
We can write this local operation on $b$ as the left
action on $S(b,a)$,
\begin{equation}\label{sec:SU2invariants:eq:phasetransformationa}
S(b,a)\rightarrow U(b) S(b,a)
\end{equation}
where the components of the new matrix $U(b)$ are given by
\begin{equation}\label{sec:SU2invariants:eq:localphaseI}
U(b)_{j_1 j_2}=\frac{1}{2}\text{tr}\left({\mathcal{U}_b}^\dag
{\sigma^b_{j_1}} \mathcal{U}_b {\sigma^b_{j_2}}\right).
\end{equation}
One can make a similar local operation, $\mathcal{U}_a$,
simultaneously on qubit $a$ and find that
\begin{equation}\label{sec:SU2invariants:eq:phasetransformation}
S(b,a)\rightarrow U(b) S(b,a) U(a)^T
\end{equation}
where
\begin{equation}\label{sec:SU2invariants:eq:localphaseII}
{U(a)^T_{i_1 i_2}}=\frac{1}{2}\text{tr}\left({\mathcal{U}_a}
{\sigma^a_{i_1}} {\mathcal{U}_a}^\dag {\sigma^a_{i_2}}\right).
\end{equation}
Under local operations we see that the
link-transformations change in the same way as the parallel
transporters in lattice gauge theories if the gauge group is an orthogonal group. The total transformation
around the closed loop described by
eq.~(\ref{sec:SU2invariants:eq:totaltrans}), under arbitrary local
operations, therefore becomes
\begin{eqnarray}\label{sec:SU2invariants:eq:totaltranslocal}
&&S(a,a;\mathcal{C})=\\ \nonumber &&U(a)S(a,z)U(z)^T \cdots U(c)S(c,b)U(b)^T
U(b)S(b,a)U(a)^T.
\end{eqnarray}
Provided the local operations cancel each other, the
trace of $S(a,a;\mathcal{C})$ is invariant under these operations. We now prove this fact specifically for $\mathcal{U} \in SU(2)$, that is $U^T U=\mathbf{I}$.
Each of the components of $U^T U$ is given by
\begin{equation}
U_{l_1 l_2}^T U_{l_2 l_3}=\frac{1}{4}\text{tr}\left({\mathcal{U}}
{\sigma_{l_1}} {\mathcal{U}}^\dag
{\sigma_{l_2}}\right)\text{tr}\left({\mathcal{U}}^\dag
{\sigma_{l_2}} \mathcal{U} {\sigma_{l_3}}\right).
\end{equation}
Writing each of the $2\times 2$ matrices in index notation we have
\begin{equation}
\frac{1}{4}\mathcal{U}_{i_1 i_2} (\sigma_{l_1})_{i_2 i_3}
{\mathcal{U}}^\dag_{i_3 i_4} (\sigma_{l_2})_{i_4 i_1}
{\mathcal{U}}^\dag_{j_1 j_2} (\sigma_{l_2})_{j_2 j_3}
\mathcal{U}_{j_3 j_4} (\sigma_{l_3})_{j_4 j_1}
\end{equation}
where in the last equation the $l$ indices take the integer values
$0 \ldots 3$ and the $i$ and $j$ indices take the integer values $0$
and $1$. Summation is implied by a repeated index. We can use the
relation
\begin{equation}\label{eq:SU2:SWAP}
\frac{1}{2}\sum_{l=0}^3(\sigma_l)_{i_1i_2}(\sigma_l)_{i_3i_4} =
\delta_{i_1i_4}\delta_{i_2i_3}
\end{equation}
($\sum_{i=0}^3 \sigma_i\otimes\sigma_i=SWAP$) and the unitary property of $\mathcal{U}$
\begin{equation}
\mathcal{U}_{i_1 i_2}{\mathcal{U}}^\dag_{i_2
i_3}={\mathcal{U}}^\dag_{i_1 i_2}\mathcal{U}_{i_2 i_3}=\delta_{i_1
i_3}
\end{equation}
to find
\begin{equation}\label{sec:SU2invariants:eq:cancellation}
U_{l_1 l_2}^T U_{l_2 l_3}=
\text{tr}\left(\sigma_{l_1}\sigma_{l_3}\right)=\delta_{l_1 l_3}.
\end{equation}
The remaining local unitary transformations, those made at the
beginning (or end) of the loop cancel from the cyclic property of
the trace. So the quantity $\text{tr }S(a,a;\mathcal{C})$ is indeed
invariant under all local unitary transformations.

A simpler way to see that the local unitary transformations do indeed
cancel is to recognize that an arbitrary unitary acting on a qubit
when written in terms of the Pauli matrices is simply a three
dimensional spatial rotation acting on the three spatial components
$\sigma_1$, $\sigma_2$ and $\sigma_3$, that is, they are just
rotations of the Bloch sphere. In other words the local operations $U(a)$,
$U(b)$ acting on qubits $a$ and $b$ respectively in the $S(b,a)$
basis can be written explicitly as
\begin{equation}
U(b)S(b,a)U(a)^T=\begin{pmatrix}
                      1 & . \\
                      . & R_b \\
                    \end{pmatrix}\begin{pmatrix}
                                   s_{00} & s_{01} & s_{02} & s_{03} \\
                                   s_{10} & s_{11} & s_{12} & s_{13} \\
                                   s_{20} & s_{21} & s_{22} & s_{23} \\
                                   s_{30} & s_{31} & s_{32} & s_{33} \\
                                 \end{pmatrix}
                    \begin{pmatrix}
                                   1 & . \\
                                   . & R_a^T \\
                                 \end{pmatrix},
\end{equation}
where $R_b$ and $R_a$ are $3\times 3$ rotation matrices, elements of
$SO(3)$. That is, in the correlation matrix basis, $\mathcal{U}\in SU(2)$ become $U\in SO(3)$ due to the well known homomorphism $SU(2)\simeq SO(3)$ \cite{ref:Arrighi&Patricot03}. The components $s_{ji}$ are expectation values of the local
spin measurements $\sigma_i^a$ and $\sigma_j^b$ made on $\rho_{ab}$. One can verify the
form of $R_a$ and $R_b$ using
eqns.~(\ref{sec:SU2invariants:eq:localphaseI}) and
(\ref{sec:SU2invariants:eq:localphaseII}).

In a similar way one can see that invariants where one performs a spin flip on each and every qubit are invariant under $\mathcal{U} \in SL(2,\mathbb{C})$ representing general local qubit operations up to a positive constant. The total transformation obtained by spin flipping every qubit can be explicitly written
\begin{eqnarray}\label{sec:SU2invariants:eq:totaltranslocalflip}
S(\tilde{a},\tilde{a};\mathcal{C})= S(a,z)\eta \cdots \eta S(c,b) \eta S(b,a)\eta.
\end{eqnarray}
Under local operations $\mathcal{U} \in SL(2,\mathbb{C})$ we have seen from equations~(\ref{sec:SU2invariants:eq:phasetransformationa}-\ref{sec:SU2invariants:eq:localphaseII}) that the correlation matrices $S$ transform as $S(b,a)\rightarrow U(b)S(b,a)U(a)^T$ thus we can form products such as $U(a)\eta U(a)^T$ from a transformation around a loop. Provided
\begin{equation}\label{sec:SU2invariants:eq:Lorentz}
U\eta U^T=\eta
\end{equation}
our spin flipped quantities $I(\tilde{a}\tilde{b}\cdots\tilde{z})$ are invariant. In fact eq.~(\ref{sec:SU2invariants:eq:Lorentz}) is the defining property of the group of Lorentz transformations, $SO^+(1,3)$, and due to the well known homomorphism $SL(2,\mathbb{C})\simeq SO^+(1,3)$ it indeed turns out that in the correlation matrix representation $\mathcal{U}\in SL(2,\mathbb{C})$ becomes $U\in SO^+(1,3)$ \cite{ref:Arrighi&Patricot03}. One can verify eq.~(\ref{sec:SU2invariants:eq:Lorentz}) holds explicitly using eqs.~(\ref{sec:SU2invariants:eq:localphaseI}) and (\ref{sec:SU2invariants:eq:localphaseII}). Therefore the spin flipped quantities $\text{tr} S(\tilde{a},\tilde{a};\mathcal{C})$ are invariant under local $SL(2,\mathbb{C})$ operations.

\subsection{Directional
property}\label{sec:SU2invariants:directional}

One other useful property of the correlation matrices or
link-transformations is simply demonstrated: The link-transformation
taking $\mathbf{m}^a$ to $\mathbf{m}^b$, the real matrix $S(b,a)$,
is the transpose of the link-transformation taking $\mathbf{m}^b$ to
$\mathbf{m}^a$. That is
\begin{equation}
S(b,a)=S(a,b)^T.
\end{equation}
This property is easily seen from
eq.~(\ref{sec:SU2invariants:eq:Sdef}). We note that this is another
property shared by the parallel transporters in lattice gauge
theory, the parallel transporter that takes you from one lattice
point to another is the transpose of the parallel
transporter that takes you back provided the gauge group is $O(N)$. However, the parallel transporters
have the additional feature that a loop not enclosing area is the
identity, for example $U(a,b)U(b,a)=\mathbf{I}$. A similar
expression for link-transformations does not hold. In fact we will
make use of this property in the following section.

\section{Identification of
invariants}\label{sec:SU2invariants:identification}

For any collection of qubits, one can consider the manifold
representing the set of {\em orbits} of pure states under all local
unitary transformations. That is, each point in the manifold
corresponds to such an orbit.  For a system of three qubits---we
call them $a$, $b$, and $c$---it is known that the manifold of
orbits is five dimensional \cite{ref:Linden&Popescu98}. (A quick but
incomplete counting argument makes this result plausible. The
eight-dimensional space of pure states can be parameterized by
fourteen real numbers, if we fix the normalization and the overall
phase. A generic orbit has nine degrees of freedom, because each of
the three local unitaries has three real parameters. This leaves
five parameters to specify the orbit itself.) Stating this in an
alternative way for the case of a pure three qubit state,
$\ket{\psi}$ is locally equivalent to $\ket{\phi}$ provided all
local invariants specifying the orbit are equal \footnote{For the
case of pure three qubit states we need five continuous local
polynomial invariants and one binary polynomial invariant to
identify which states are locally equivalent. This will be discussed
later in the section.}. In the case of equality $\ket{\psi}$ and
$\ket{\phi}$ have the same entanglement properties and one can
obtain $\ket{\phi}$ from $\ket{\psi}$ simply by making local unitary transformations on
each qubit.

Several authors have studied local invariants of pure three qubit
states
\cite{ref:Coffman00,ref:Acin00,ref:Carteret00,ref:Linden&Popescu98,ref:Barnum01,ref:Gingrich02,ref:Leifer04,ref:Brun01,ref:Sudbery01}.
In particular, Sudbery \cite{ref:Sudbery01} has identified a
convenient set of algebraically independent invariants, each of
which is a polynomial in the eight complex components of the vector
$|\psi\rangle$.  Not counting the normalization (which is Sudbery's
$I_1$), there are five of these invariants, the same as the number
of dimensions:
\begin{eqnarray}\label{sec:SU2invariants:eq:Sudberys}
& I_2 = \hbox{tr}[(\rho_c)^2], \; \;
 I_3 = \text{tr}[(\rho_b)^2] , \;  \;
 I_4 = \text{tr}[(\rho_a)^2]   \nonumber \\
& I_5 = (\rho_{ab})_{i j',i' j}\,(\rho_{bc})_{j k',j' k}\,(\rho_{ca})_{k i',k' i}  \nonumber \\
& I_6 = (1/4)\tau_{abc}^2
\end{eqnarray}
where summation over repeated indices is implied in the definition
of $I_5$.  Here each index takes the values 0 and 1, and we have
used the letters $i$, $j$, and $k$ to refer to qubits $a$, $b$, and
$c$ respectively.  The Kempe invariant $I_5$ \cite{ref:Kempe99} can
be written in several different ways, the above form being most
convenient for our purpose. The quantity $\tau_{abc}$, is the 3-tangle which
measures a kind of three-way entanglement characteristic of the GHZ
state $(1/\sqrt{2})(|000\rangle + |111\rangle)$
\cite{ref:Coffman00}.  If we write $|\psi\rangle$ in terms of the
standard basis states as $|\psi\rangle = \sum a_{ijk}|ijk\rangle$,
then the invariant $I_6$ can be expressed as
\begin{equation}
I_6=\left|
\epsilon_{i_1i_2}\epsilon_{i_3i_4}\epsilon_{j_1j_2}\epsilon_{j_3j_4}
\epsilon_{k_1k_3}\epsilon_{k_2k_4}a_{i_1j_1k_1}a_{i_2j_2k_2}a_{i_3j_3k_3}a_{i_4j_4k_4}\right|^2,
\end{equation}
where $\epsilon_{ij}$ is the antisymmetric tensor in two dimensions.

The invariants listed in eq.~(\ref{sec:SU2invariants:eq:Sudberys})
are not complete in the sense of determining a unique orbit.  In
particular, these invariants do not distinguish between a state and
its complex conjugate, which may well lie on different orbits.
Because $I_1$ - $I_6$ are real
\begin{equation}
I_i(\ket{\psi})=I_i(\ket{\psi}^*).
\end{equation}
As reported by Ac\`{i}n \emph{et al.} \cite{ref:Acin01}, Grassl has
shown that this ambiguity can be removed by including a single
binary invariant based on a complex twelfth-degree polynomial in the amplitudes $a_{ijk}$.

We now ask whether the invariants in
eq.~(\ref{sec:SU2invariants:eq:Sudberys}) can be generated via the
formalism of section~\ref{sec:SU2invariants:path}. The first three
can indeed be expressed quite simply in this way. For example,
\begin{eqnarray}
I(ab) &=& \text{tr}\left\{ S(a,b)S(b,a)\right\} \nonumber  \\
&=& \sum_{ij} \text{tr}\left[\left( \frac{\sigma_i^a\otimes\sigma_j^b}{2}\right)\rho_{ab}\right]^2 \nonumber\\
&=& \text{tr}[(\rho_{ab})^2] = \text{tr}[(\rho_c)^2] = I_2.
\end{eqnarray}
The last line follows from the fact that the operators
$(\sigma_i^a\otimes\sigma_j^a)/2$ constitute a complete orthonormal
basis for the space of $4\times 4$ matrices.

The Kempe invariant $I_5$ fits particularly well into our scheme.
As we now show, this invariant is simply
\begin{equation}
I_5 = I(abc)=\text{tr}\left\{S(a,c)S(c,b)S(b,a)\right\}.
\end{equation}
To see this, we start with the following expression for $I(abc)$:
\begin{equation}
\frac{1}{8}\text{tr
}\hspace{-1mm}\left[\left(\sigma^a_l\otimes\sigma^b_m\right)\hspace{-1mm}\rho_{ab}\right]
\text{tr
}\hspace{-1mm}\left[\left(\sigma^b_m\otimes\sigma^c_n\right)\hspace{-1mm}\rho_{bc}\right]
\text{tr
}\hspace{-1mm}\left[\left(\sigma^c_n\otimes\sigma^a_l\right)\hspace{-1mm}\rho_{ca}\right]
\end{equation}
in which summation over $l$, $m$, and $n$ is implied.  This
summation considerably simplifies the expression, because of eq.~(\ref{eq:SU2:SWAP}). This relation tells us how to connect up the indices of the three
density matrices, and we obtain the interlocking pattern that we saw
in eq.~(\ref{sec:SU2invariants:eq:Sudberys}):
\begin{equation}\label{sec:SU2invariants:eq:Kempe}
I(abc) = (\rho_{ab})_{i j',i' j}\,(\rho_{bc})_{j k',j'
k}\,(\rho_{ca})_{k i',k'i} = I_5.
\end{equation}

\begin{figure}
\begin{center}
\subfigure[$I(ab)$]{\includegraphics[width=2cm]{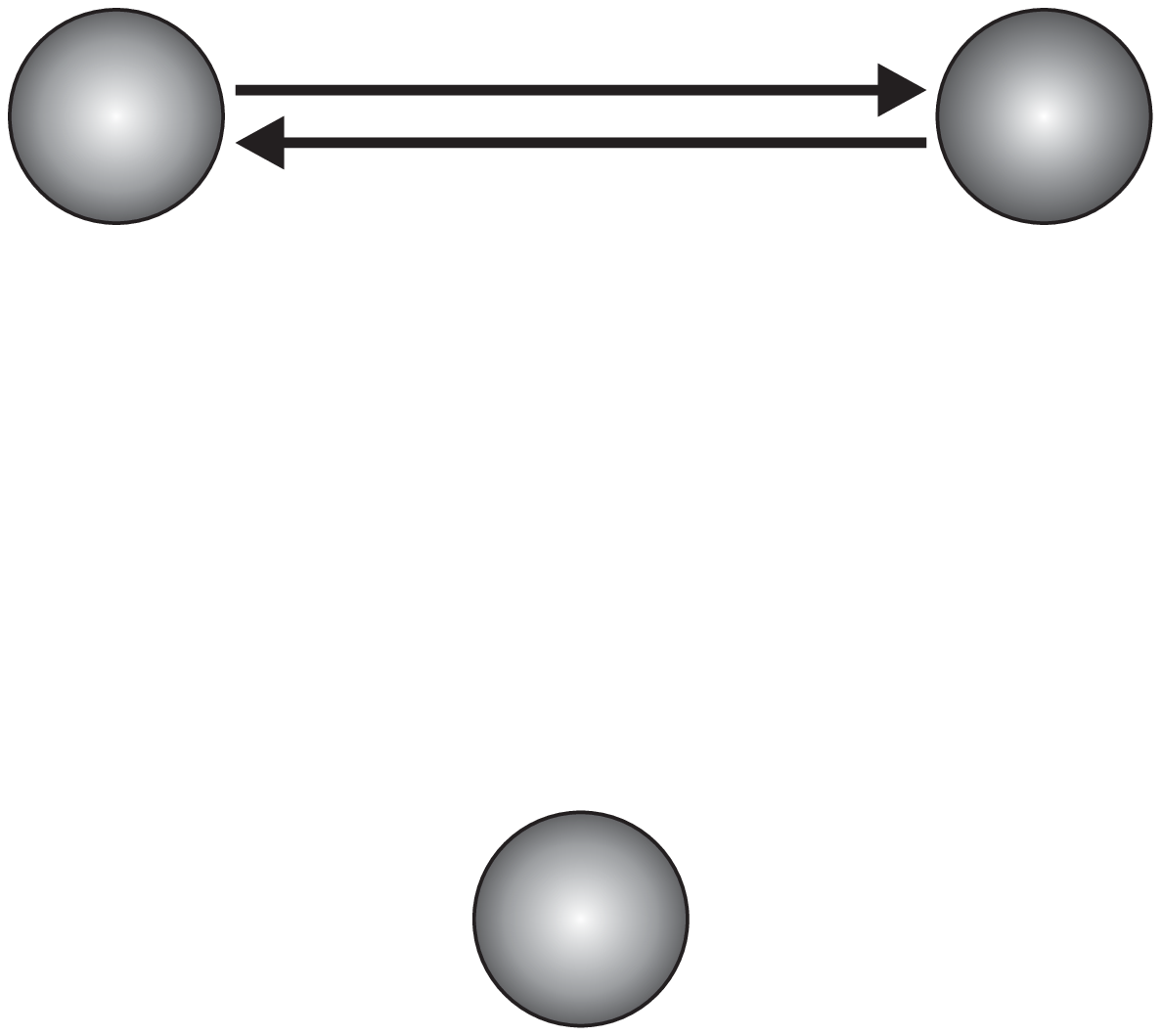}\label{sec:SU2invariants:fig:I2}}
\hspace{2cm}
\subfigure[$I(bc)$]{\includegraphics[width=2cm]{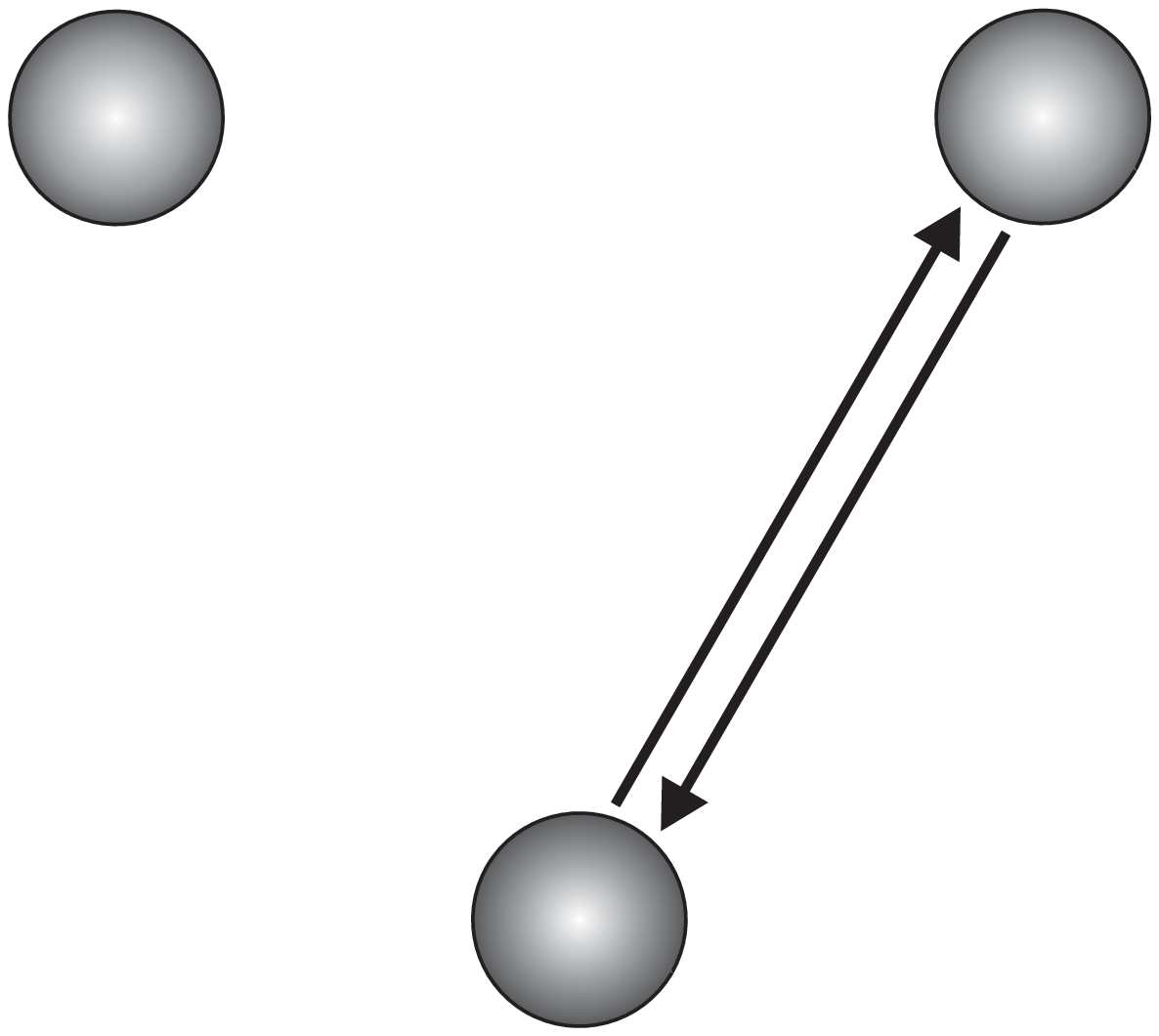}\label{sec:SU2invariants:fig:I3}}
\hspace{2cm}
\subfigure[$I(ca)$]{\includegraphics[width=2cm]{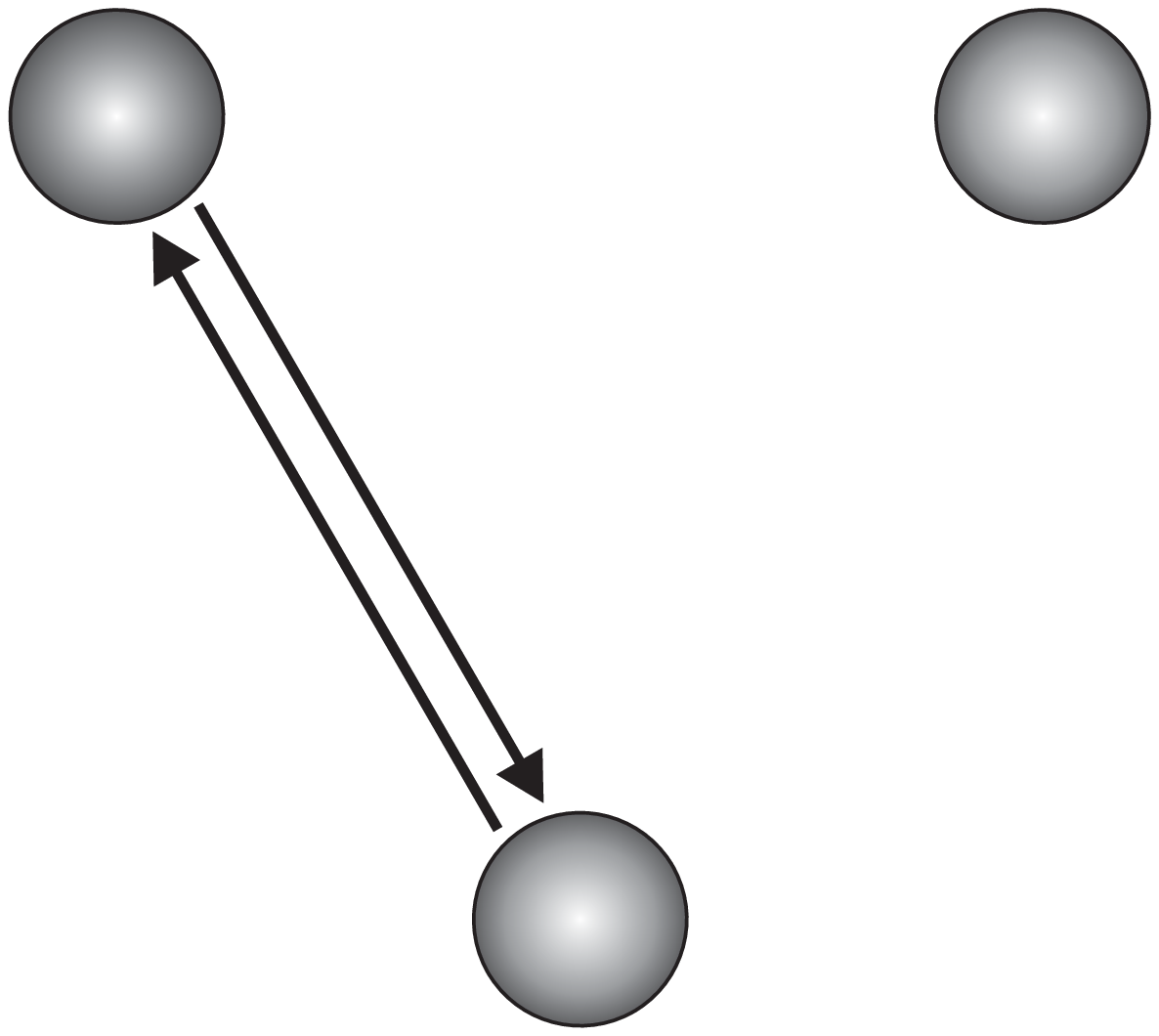}\label{sec:SU2invariants:fig:I4}}\\
\subfigure[$I(abc)$]{\includegraphics[width=2cm]{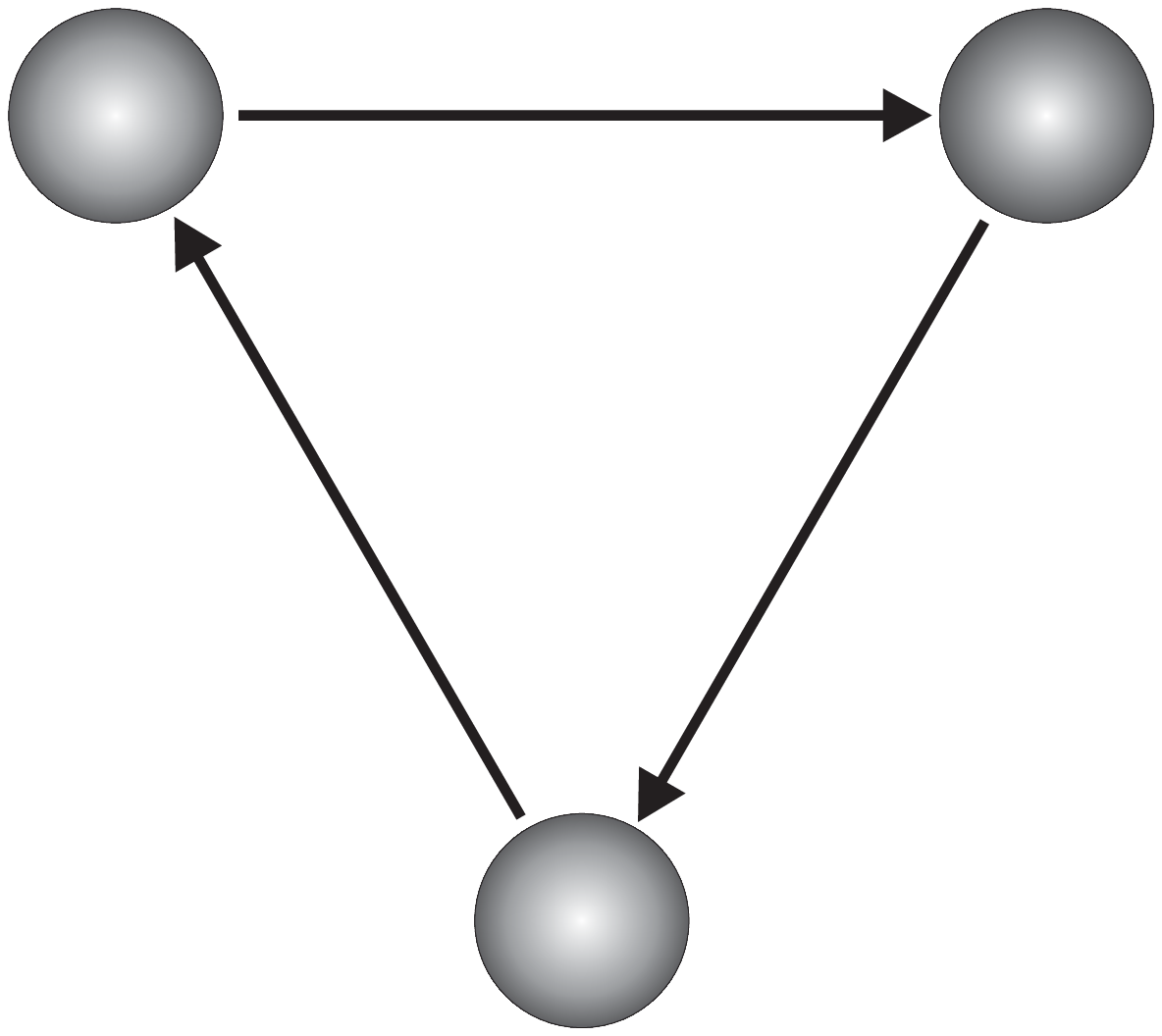}\label{sec:SU2invariants:fig:I5}}
\hspace{2cm}
\subfigure[$I(abab)$]{\includegraphics[width=2cm]{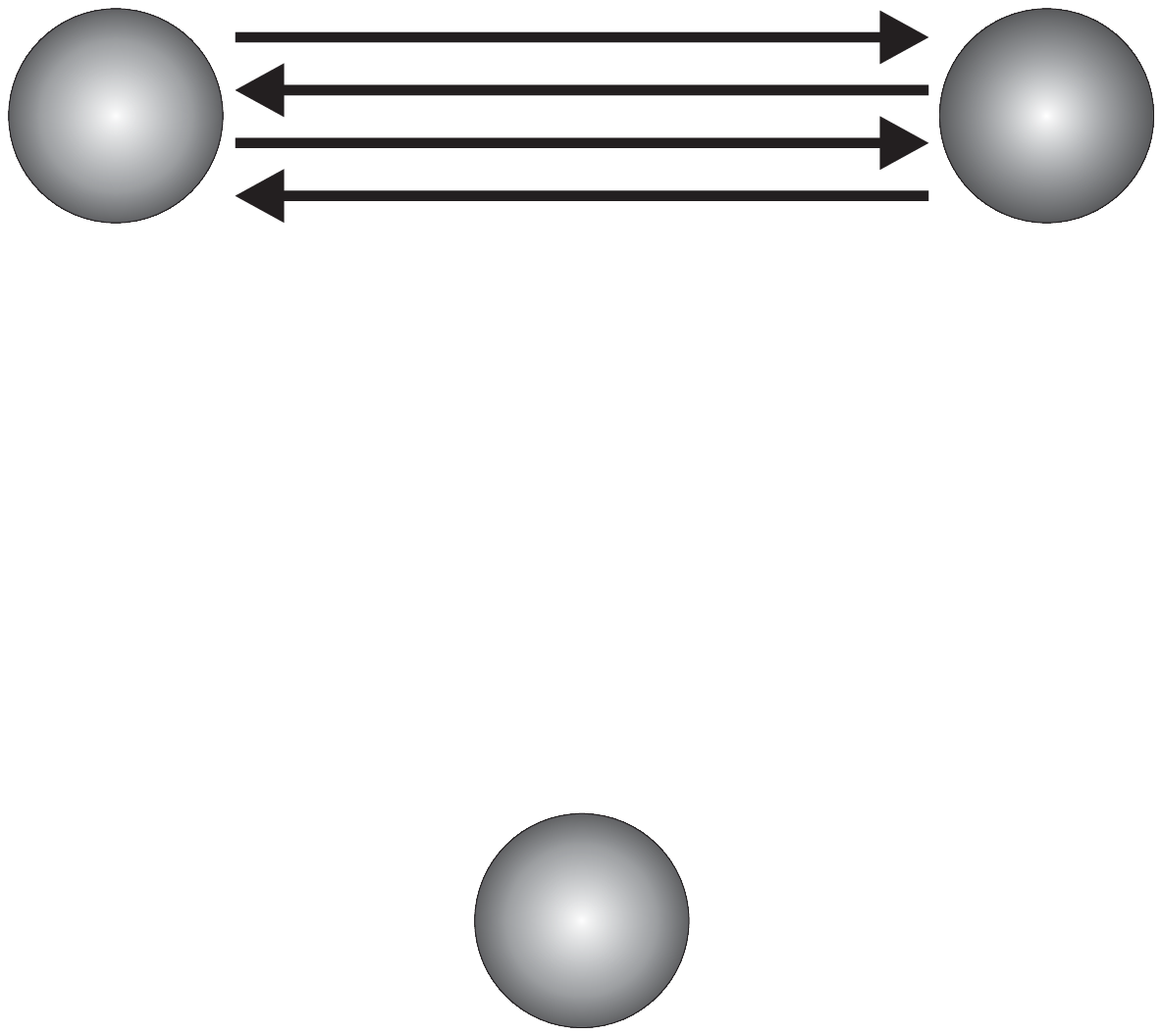}\label{sec:SU2invariants:fig:I6}}
\end{center}
\caption[Closed paths giving the set of $SU(2)$ local invariants for
pure states of three qubits.]{Graphical representations of the
closed paths giving the set of polynomial local $SU(2)$ invariants
for pure states of three qubits.}\label{sec:SU2invariants:fig:invariantpaths}
\end{figure}

Of the set of invariants that Sudbery identifies, the only one
remaining is $I_6$, a degree-eight polynomial in the components
$a_{ijk}$ and their conjugates which is proportional to the square
of the 3-tangle. $I_6$ is also invariant under $SL(2,\mathbb{C})$
unlike $I_1$ - $I_5$ which do not have this higher invariance. Our
formalism does not produce $I_6$ directly, though we can easily
generate a different polynomial of degree eight that is likewise
algebraically independent of the first four invariants. It is
defined by any path that cycles twice between two of the qubits,
that is, any of the invariants
\begin{equation}
I(abab) = I(bcbc) = I(caca).
\end{equation}
One can write down this invariant directly in terms of the reduced
density matrices $\rho_{ab}$ and $\rho_{ba}$ (which are related to
each other by the swap operation that interchanges the two qubits),
following precisely the pattern of index connections that we see in
eq.~(\ref{sec:SU2invariants:eq:Kempe}). Now, however, because the
same two-step path is repeated, we use subscripts 1 and 2 on the
indices to distinguish the two round-trips.
\begin{eqnarray}
&&I(abab)
=\text{tr}\left\{S(a,b)S(b,a)S(a,b)S(b,a)\right\}\nonumber\\
&=&(\rho_{ab})_{i_1j'_1,i'_1j_1}(\rho_{ba})_{j_1i'_2,j'_1i_2}
(\rho_{ab})_{i_2j'_2,i'_2j_2}(\rho_{ba})_{j_2i'_1,j'_2i_1}.
\end{eqnarray}
We can alternatively write out this invariant in terms of the
components $a_{ijk}$:
\begin{eqnarray}
I(abab) &=& a_{i_1j_2k_1} a^*_{i_2j_1k_1}a_{i_4j_1k_2} a^*_{i_3j_2k_2}\nonumber  \\
&\times&  a_{i_3j_4k_3} a^*_{i_4j_3k_3}a_{i_2j_3k_4}
a^*_{i_1j_4k_4}.
\end{eqnarray}
In this latter form it is clear that the invariant is symmetric
under permutations of the qubits: by permuting the factors of $a$
and $a^*$, one can interchange the roles of the $i$, $j$, and $k$
indices.

To show that the invariants we have identified are algebraically
independent, it is sufficient to show that their gradients at any
point, together with the gradient of the normalization invariant
$I_1=a_{ijk} a^*_{ijk}$, constitute a linearly independent set of
vectors \cite{ref:Sudbery01}. One finds that this is indeed the
case. So we now have the following list of path-based
invariants, not quite identical to Sudbery's but no less complete:
\begin{eqnarray}
I(ab)&=&\text{tr}\left\{S(a,b)S(b,a)\right\}, \nonumber\\
I(bc)&=&\text{tr}\left\{S(b,c)S(c,b)\right\}, \nonumber\\
I(ca)&=&\text{tr}\left\{S(c,a)S(a,c)\right\}, \nonumber\\
I(abc)&=&\text{tr}\left\{S(a,c)S(c,b)S(b,a)\right\}, \nonumber\\
I(abab)&=&\text{tr}\left\{S(a,b)S(b,a)S(a,b)S(b,a)\right\},
\end{eqnarray}
the last one being symmetric under permutations of the three qubits
even though the path it is based on is not. The three kinds of path
we have used in constructing our invariants are illustrated in
figure~\ref{sec:SU2invariants:fig:invariantpaths}.


Notice that our construction provides an interpretation of
the Kempe invariant $I(abc) = I_5$. In our measurement scenario, in
which each successive measurement collapses the state of the next
qubit, the Kempe invariant is the trace of the transformation that
results from following the triangular path through all three qubits.
Recall that at each stage in this measurement scenario, the trace of
the new $M$ matrix is equal to the probability of getting the
desired outcome. Thus, the resulting invariant tends to be larger if
the collapsed state at each step is strongly represented in the
original reduced density matrix of the qubit in question.  The most
extreme example of this kind of consistency is the case of a
completely factorable state, in which the collapsed state {\em must}
be proportional to the original pure state of the given qubit.  And
indeed, the Kempe invariant is largest when the state is fully
factorable ($I(abc)=1$). One can also show the Kempe invariant takes its minimal value for the W state $\ket{001}+\ket{010}+\ket{100}$ at $I(abc)=2/9$ \cite{ref:Osterloh08}.

\section{Local invariants using the spin flip}\label{sec:SU2invariants:SL2}

In section~\ref{sec:SU2invariants:identification} we provided a complete set of
algebraically independent local $SU(2)$ invariants by considering
different closed paths around the lattice. In this section we
again consider closed paths around the lattice but this time
including the spin flip operation on every qubit lifting the
invariance of the quantities produced to $SL(2,\mathbb{C})$.

From the $SU(2)$ invariant list, it turns out that we can replace
$I(ab)$, $I(bc)$, and $I(ca)$ with $I(\tilde{a}\tilde{b})$,
$I(\tilde{b}\tilde{c})$, and $I(\tilde{c}\tilde{a})$; the invariants
are still independent. Moreover, these `flipped' invariants can also
be interpreted in terms of entanglement. One finds that
\begin{eqnarray}\label{sec:SL2invariants:eq:quartic}
I(\tilde{a}\tilde{b}) =
\text{tr}\left[\rho_{ab}\tilde{\rho}_{ab}\right] = \tau_{ab} +
\frac{1}{2}\tau_{abc}\nonumber\\
I(\tilde{b}\tilde{c}) =
\text{tr}\left[\rho_{bc}\tilde{\rho}_{bc}\right] = \tau_{bc} +
\frac{1}{2}\tau_{abc}\nonumber\\
I(\tilde{c}\tilde{a}) =
\text{tr}\left[\rho_{ca}\tilde{\rho}_{ca}\right] = \tau_{ca} +
\frac{1}{2}\tau_{abc}
\end{eqnarray}
Here $\tilde{\rho} =
(\sigma_2\otimes\sigma_2)\rho^T(\sigma_2\otimes\sigma_2)$ is the
spin flipped state of the two qubits, and $\tau_{ab}$ is the tangle
between qubits $a$ and $b$, a measure of their pairwise entanglement (it is the square of the concurrence)
\cite{ref:Wootters98}. The completely flipped version of $I(abab)$,
that is, $I(\tilde{a}\tilde{b}\tilde{a}\tilde{b})$, likewise
produces a non-trivial invariant but it is not algebraically
independent of $I(\tilde{a}\tilde{b})$, $I(\tilde{b}\tilde{c})$, and
$I(\tilde{c}\tilde{a})$. A natural eighth order $SL(2,\mathbb{C})$
invariant is given by the determinant of any of the
link-transformations, for example:
\begin{eqnarray}
\det\left[S(a,b)\right]=-\frac{1}{16}\tau_{ab}(\tau_{ab}+\tau_{abc}).
\end{eqnarray}
One can see the determinant of the link-transformations is indeed
$SL(2,\mathbb{C})$ invariant from the way the link-transformations
change under arbitrary local $SL(2,\mathbb{C})$ transformations
(eq.~(\ref{sec:SU2invariants:eq:phasetransformation})) and the property
of the determinant $\det(AB)=\det(A)\det(B)$. We can show this
invariant is algebraically independent of $I(\tilde{a}\tilde{b})$,
$I(\tilde{b}\tilde{c})$, and $I(\tilde{c}\tilde{a})$ using the same
methods of section~\ref{sec:SU2invariants:identification}.

So far we have a set of four $SL(2,\mathbb{C})$ invariants. They tell us about the entanglements in the
state since one can reconstruct the \emph{amounts} of entanglement,
$\tau_{ab}$, $\tau_{bc}$, $\tau_{ac}$ and $\tau_{abc}$ from just
these four invariants. For example, the 3-tangle can be expressed as
\begin{equation}
\tau_{abc} = 2\sqrt{16\det S(a,b) + I(\tilde{a}\tilde{b})^2}.
\end{equation}
Even though the three qubit labels do not enter this expression
symmetrically (there is no explicit reference to qubit $c$), the
3-tangle is symmetric under permutations of the qubits. Similar
expressions can be written for the 2-tangles.

These four amounts of entanglement do not form a complete set of
algebraically independent invariants. To complete this set we could
use the $SU(2)$ Kempe invariant which one can verify is
algebraically independent of the four tangles, however we would also
like an invariant with $SL(2,\mathbb{C})$ local invariance. A
natural choice of loop is the one that gave the Kempe invariant.
Strangely enough, the completely flipped version of the Kempe
invariant, that is, $I(\tilde{a}\tilde{b}\tilde{c})$, turns out to
be exactly zero for all pure three qubit states as we prove in the
following subsection.

\begin{figure}
\begin{center}
\subfigure[$I(\tilde{a}\tilde{b})$]{\includegraphics[width=2cm]{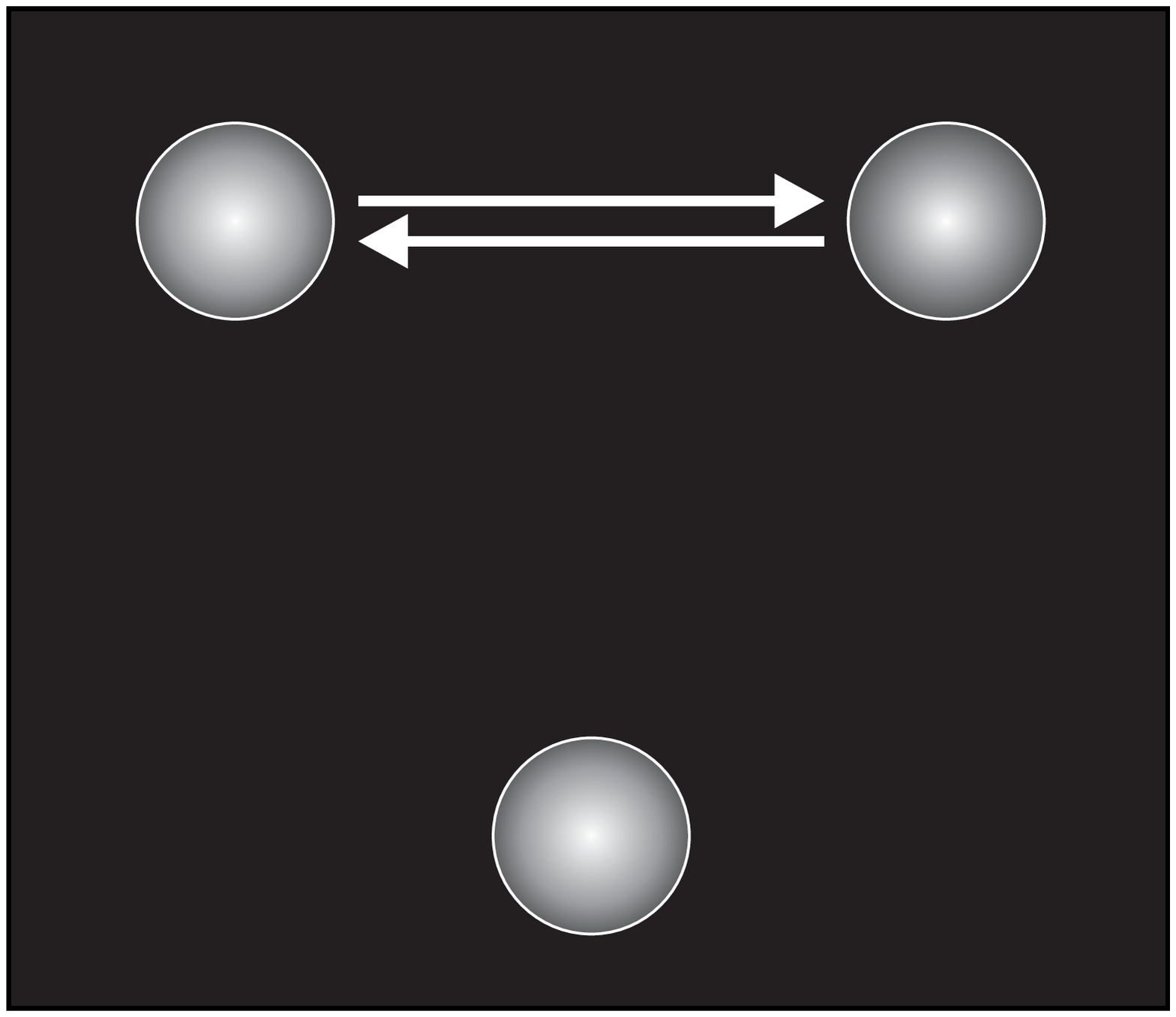}\label{sec:SL2invariants:fig:I2}}
\hspace{2cm}
\subfigure[$I(\tilde{b}\tilde{c})$]{\includegraphics[width=2cm]{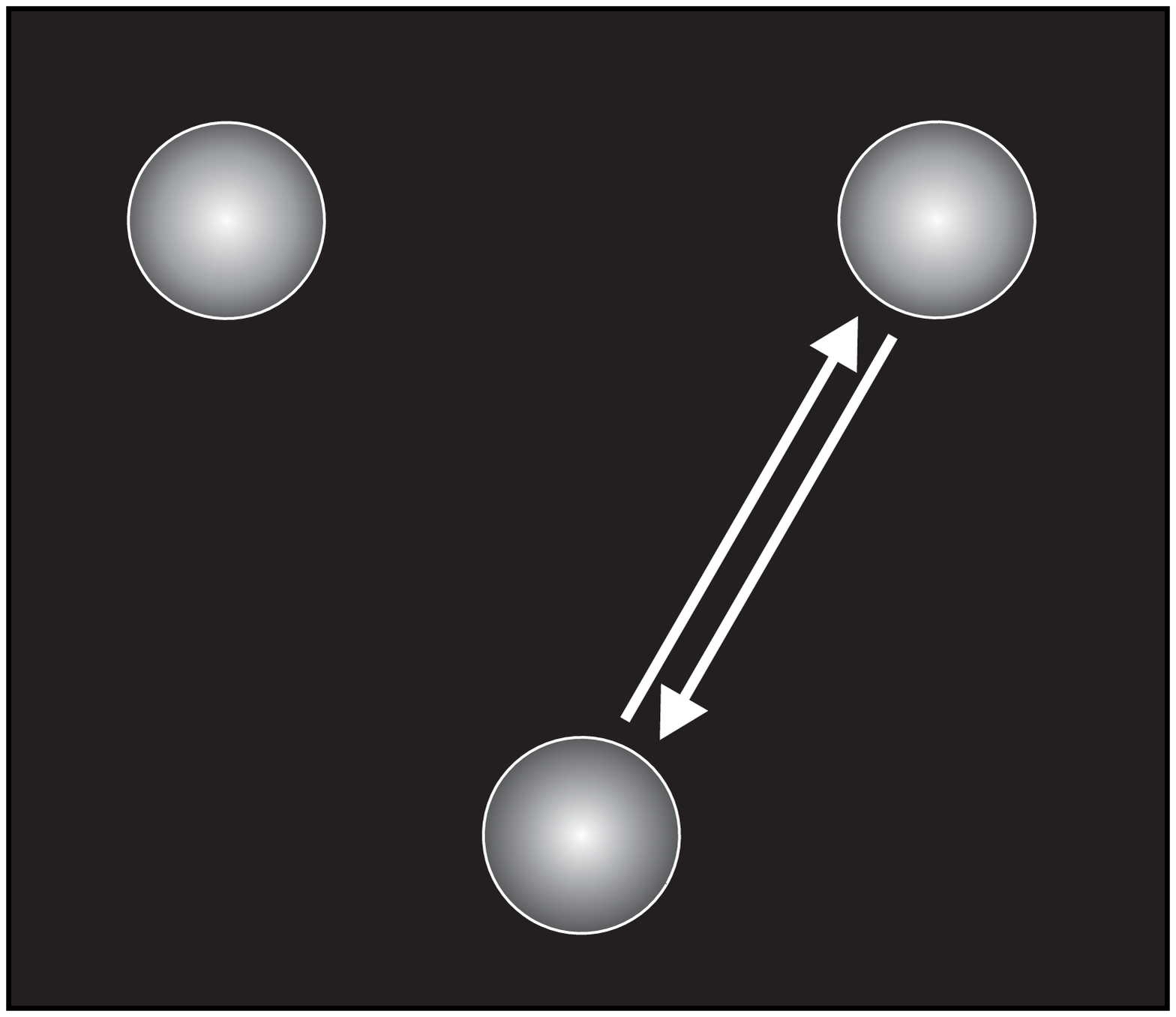}\label{sec:SL2invariants:fig:I3}}
\hspace{2cm}
\subfigure[$I(\tilde{c}\tilde{a})$]{\includegraphics[width=2cm]{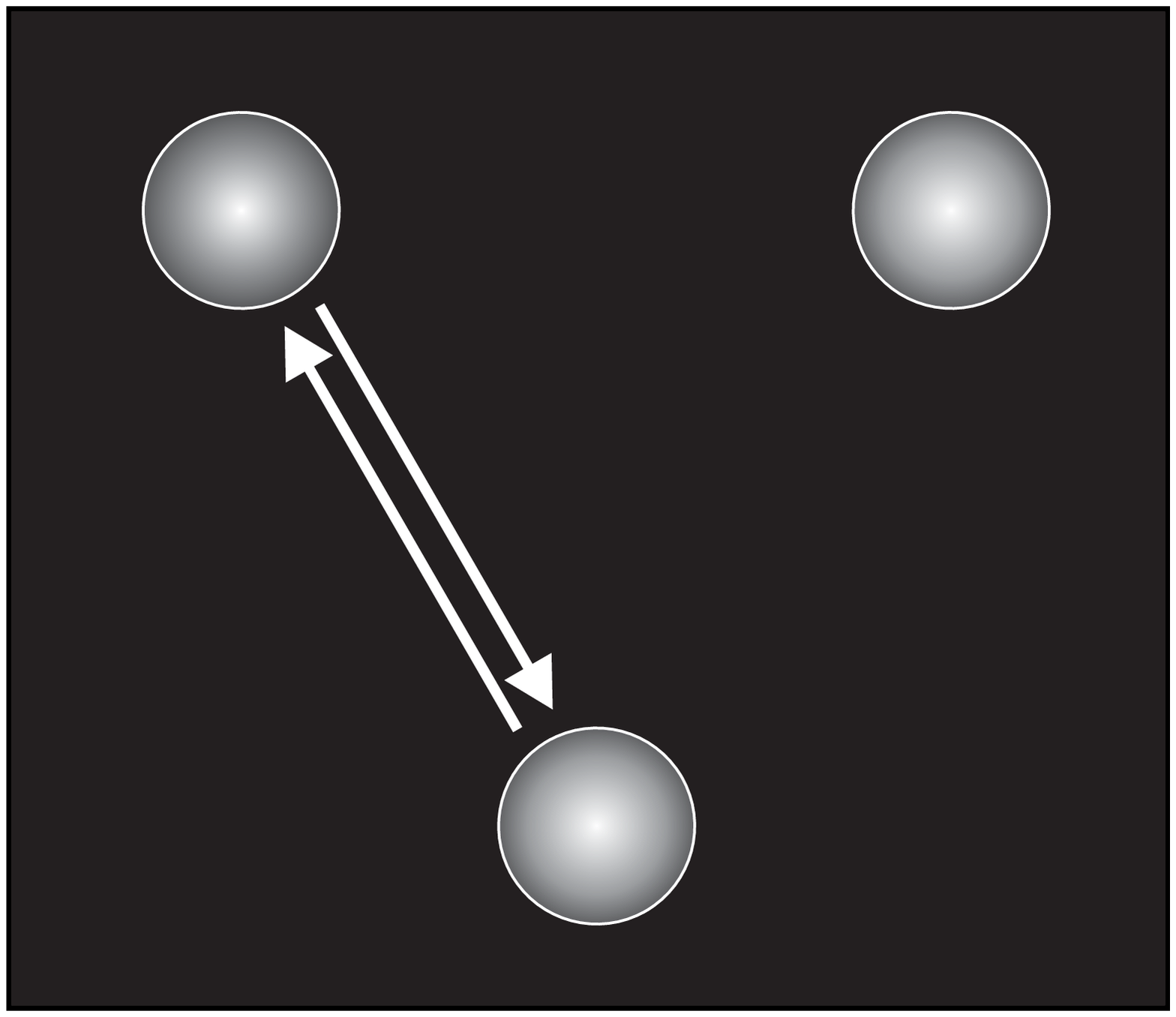}\label{sec:SL2invariants:fig:I4}}\\
\subfigure[$I(\tilde{a}\tilde{b}\tilde{c})$]{\includegraphics[width=2cm]{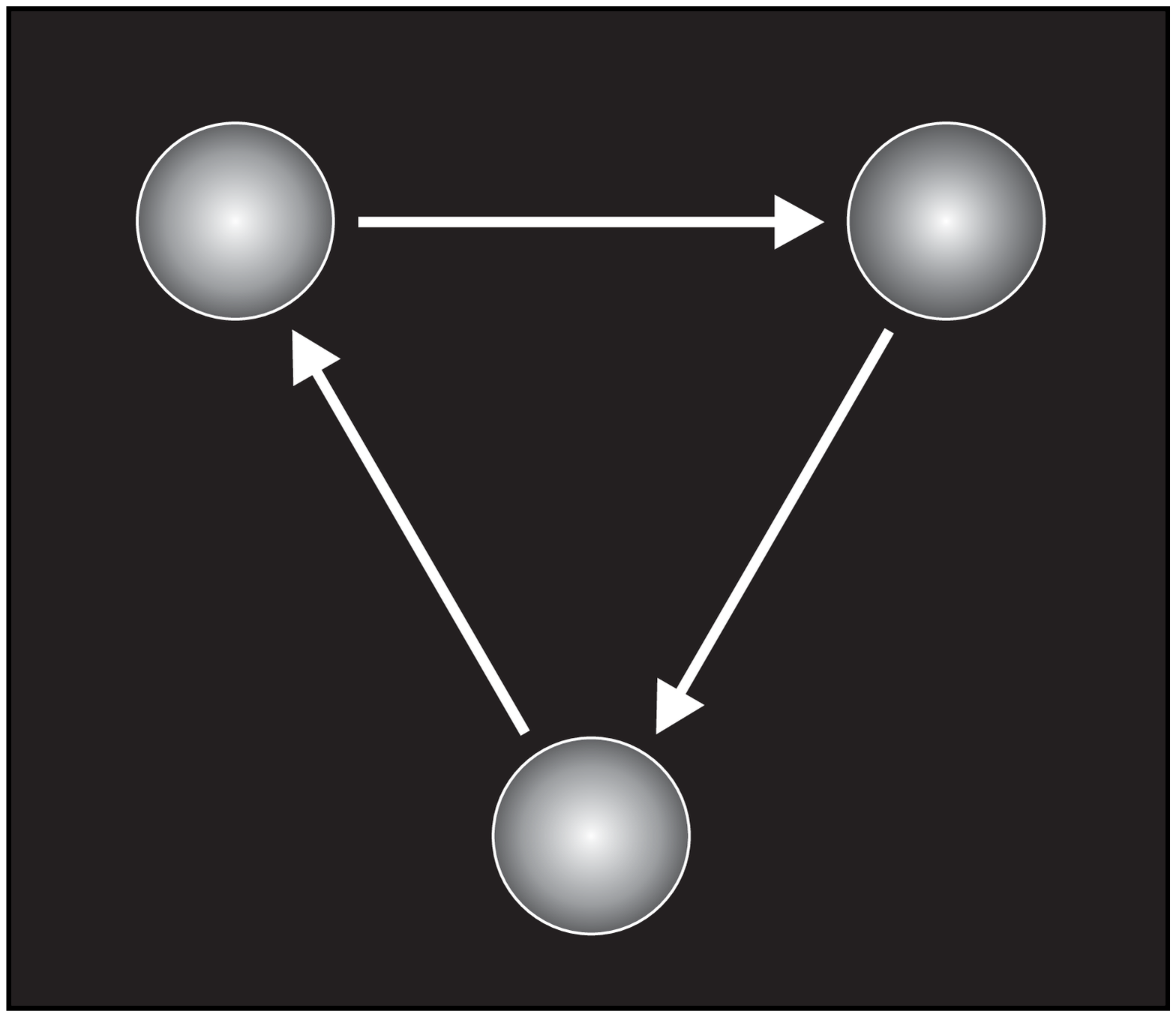}\label{sec:SL2invariants:fig:I5}}
\hspace{2cm} \subfigure[$\det
(S(a,b))$]{\includegraphics[width=2cm]{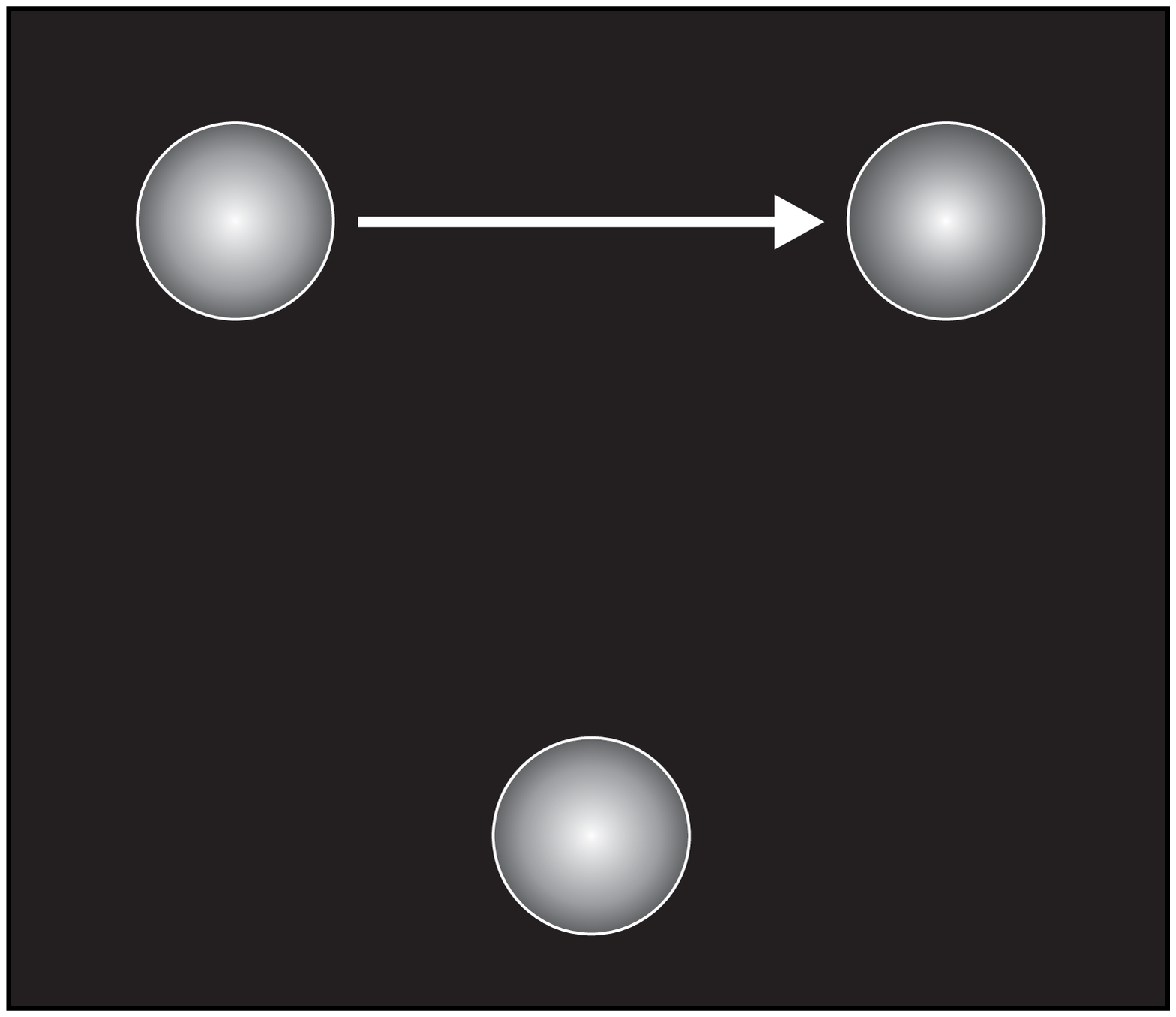}\label{sec:SL2invariants:fig:I6}}
\end{center}
\caption[Closed paths giving the set of $SL(2,\mathbb{C})$ local
invariants for pure states of three qubits.]{Graphical
representations of the closed paths giving the set of polynomial
local $SL(2,\mathbb{C})$ invariants for pure states of three qubits.
Notice that $I(\tilde{a}\tilde{b}\tilde{c})$, the flipped Kempe
invariant, is the only closed path to enclose area. It is also
always equal to zero.}\label{sec:SL2invariants:fig:invariantpaths}
\end{figure}

\subsection{Proof of
$I(\tilde{a}\tilde{b}\tilde{c}) = 0$ for any pure three qubit
state.}

From the definition of $I(\tilde{a}\tilde{b}\tilde{c})$, one finds
directly that
\begin{equation}
\begin{split}
I(\tilde{a}\tilde{b}\tilde{c}) =
&1 - \text{tr}(\rho_a^2) - \text{tr}(\rho_b^2) -\text{tr}(\rho_c^2) \\
&+\text{tr}[(\rho_a\otimes \rho_b)\rho_{ab}]
+ \text{tr}[(\rho_b\otimes \rho_c)\rho_{bc}] \\
&+\text{tr}[(\rho_a\otimes \rho_c)\rho_{ac}] - I_5.
\end{split}
\end{equation}
Sudbery \cite{ref:Sudbery01} showed that Kempe's invariant can also
be written as
\begin{equation}
\begin{split}
I_5&=3\text{tr}[(\rho_a\otimes\rho_b)\rho_{ab}]-\text{tr}(\rho_a^3)-\text{tr}(\rho_b^3)\\
&=3\text{tr}[(\rho_b\otimes\rho_c)\rho_{bc}]-\text{tr}(\rho_b^3)-\text{tr}(\rho_c^3)\\
&=3\text{tr}[(\rho_a\otimes\rho_c)\rho_{ac}]-\text{tr}(\rho_a^3)-\text{tr}(\rho_c^3).
\end{split}
\end{equation}
Using these relations we can rewrite
$I(\tilde{a}\tilde{b}\tilde{c})$ as
\begin{equation}
\begin{split}\label{sec:SL2invariants:eq:flippedKempe}
I(\tilde{a}\tilde{b}\tilde{c})
=&1-[\text{tr}(\rho_a^2)+\text{tr}(\rho_b^2)+\text{tr}(\rho_c^2)]\\
+&\frac{2}{3}[\text{tr}(\rho_a^3)+\text{tr}(\rho_b^3)+\text{tr}(\rho_c^3)].
\end{split}
\end{equation}
This last expression is a function only of the trace of powers of
the single-qubit density matrices. The Cayley-Hamilton theorem for
any $2 \times 2$ matrix $X$ is
\begin{equation}
X^2-\text{tr}(X)X+\det(X)\mathbf{I}=0.
\end{equation}
Multiplying this expression by $X$, taking the trace, and then using
the fact that for a single qubit, $\det \rho = (1/2)(1 -
\text{tr}\rho^2)$, we obtain the relation
\begin{equation}
1/3 - \text{tr}(\rho^2) + (2/3)\text{tr}(\rho^3) = 0,
\end{equation}
which together with eq.~(\ref{sec:SL2invariants:eq:flippedKempe})
shows that $I(\tilde{a}\tilde{b}\tilde{c}) = 0$.

\section{Operational interpretation of invariants}\label{sec:SU2invariants:interpretation}

Going back to our original thought experiment in
section~\ref{sec:SU2invariants:path} we can obtain a rigorous
operational interpretation of our invariants as follows if we drop the restriction that $M_a$ is positive but still Hermitian. In other words $M_a$ represents an observable rather than a measurement outcome. If we relax the positivity our
invariant can be thought of as the average fidelity between our
initial observable outcome $M_a$ and our final observable outcome
$M_a'$ resulting from the transform around the loop associated to
the particular invariant, $I(\mathcal{C})$. The average is taken
over all possible initial outcomes $M_a$ with a fixed size.
Making this idea more precise, we have
\begin{equation}
I(\mathcal{C})\propto \langle \text{tr}(M_a^\dag M_a') \rangle,
\end{equation}
where the brackets $\langle . \rangle$ denote the average.

The constraint on $M_a$'s size is given by
\begin{equation}\label{sec:SU2invariants:eq:constraint}
\text{tr} (M_a^\dag M_a)=2 k^2.
\end{equation}
In terms of the Pauli operator basis, this condition can be
written as
\begin{equation}
\text{tr} (M_a^\dag M_a)=2 \sum_{i=0}^3 (m_i^a)^2=2 k^2.
\end{equation}
The $m_i^a$ must be real for $M_a$ to be Hermitian and represent the
outcome of an observable. Similarly we can write the outcome on $a$
following the loop $\mathcal{C}$ in the Pauli operator basis as
(dropping the sub- and superscript $a$)
\begin{equation}
M'=\sum_{i=0}^3 m_i' \sigma_i
\end{equation}
where
\begin{equation}
m_i'=S(\mathcal{C})_{ij} m_j.
\end{equation}
$S(\mathcal{C})$ is the total transformation around the loop.

We can substitute these expressions into the equation for the
fidelity between transformed and initial observable outcomes to give
\begin{eqnarray}
\text{tr} (M^\dag M')&=&2\sum_{i=0}^3 m_i m_i' \nonumber\\
&=&2\sum_{i,j=0}^3 m_i S(\mathcal{C})_{ij}m_j.
\end{eqnarray}
Our invariant $I(\mathcal{C})$ is given by the elements
$S(\mathcal{C})_{ii}$ and therefore we want to find an expression
solely in terms of these elements.

We now average this fidelity. Since
eq.~(\ref{sec:SU2invariants:eq:constraint}) is the equation of a
3-sphere with radius $k$ we can perform the average over the surface
of the 3-sphere. Writing the $m_i$ in hyper-spherical coordinates we
have
\begin{eqnarray}
m_0&=&k\cos\phi_1\nonumber\\
m_1&=&k\sin\phi_1\cos\phi_2\nonumber\\
m_2&=&k\sin\phi_1\sin\phi_2\cos\phi_3\nonumber\\
m_3&=&k\sin\phi_1\sin\phi_2\sin\phi_3.
\end{eqnarray}
We can now compute the average fidelity in terms of these
coordinates. It is given by
\begin{eqnarray}
\langle \text{tr} \left(M^\dag M'\right)\rangle =
\frac{2}{A}\sum_{i,j=0}^3 S(\mathcal{C})_{ij}\int_\mathcal{S}  m_i
m_j d\mathcal{S}.
\end{eqnarray}
$\mathcal{S}$ is the entire surface of the 3-sphere,
$d\mathcal{S}=k^2 \sin^2\phi_1\sin\phi_2 d\phi_1 d\phi_2 d\phi_3$ is the area element and $A=2\pi^2 k^2$ is its total surface area.
One finds that
\begin{equation}
\frac{2}{A}\int_\mathcal{S}  m_i m_j d\mathcal{S}=
\frac{k^2}{2}\delta_{ij}
\end{equation}
and we obtain our desired result
\begin{equation}
\langle \text{tr} (M^\dag M')\rangle=\frac{k^2}{2} I(\mathcal{C}).
\end{equation}

One can also average over all observable sizes (or strengths) $k$ to obtain the same result up to a constant. The surface integral over the sphere now becomes a volume integral over the 3-ball. We also note we can choose $M_a$ to be an element of $SU(2)$. That is, a unitary that does not have to be Hermitian. In this case $k=1$ and the average is over the three Euler angles describing a element of this group. For $SU(2)$ $m_0$ is real and $m_1$, $m_2$, $m_3$ are purely imaginary. The proof goes through in the same way.

\section{Conclusions}\label{sec:SU2invariants:conclusion}

In this paper we have presented a geometric approach to
constructing quantities that are invariant under local $SU(2)$ and $SL(2,\mathbb{C})$
transformations. Our basic construction corresponds to a scenario in
which a measurement outcome on each particle along a closed path
defines the state of the next particle. We have seen that one can
produce in this way an algebraically independent set of five
invariants for a pure state of three qubits, almost identical to the
set of invariants identified by Sudbery.  One of these quantities,
the Kempe invariant, has been difficult to interpret as an amount of
entanglement.  In our construction, though, it is the one that
emerges the most naturally.  Unlike the other four invariants, the
Kempe invariant $I_5 = I(abc)$ corresponds to a path that `encloses
area' in the sense that one does not retrace one's steps. This
property sets the Kempe invariant apart from the others. Notice
that for an area enclosing path, one needs at least three qubits. In a future paper
we will exploit this area enclosing property and the existence of
a special form of a polar decomposition for correlation matrices to find
quantities much more analogous to lattice gauge field theories. The gauge
group will turn out to be the group of Lorentz transformations and
has an operational interpretation in terms of general local qubit operations. The invariants, the Wilson loops, in this construction, will be related to the curvature of the correlation space \cite{ref:Williamson10b}.

We have also provided an operational interpretation of the
invariants, including the Kempe invariant, in terms of the average
fidelity between initial and transformed observable outcomes.

We have concentrated on pure three qubit states as a test ground for
our ideas, however there is nothing specific here about the numbers
of qubits of our quantum state. Neither is there any requirement for
the state to be pure or for the subsystems to be two level. The framework presented here can be applied to
any qu$d$it state to generate local unitary invariants. To make this generalization, one replaces the Pauli matrices specific for qubits, by the generalized Gell-Mann matrices $\lambda_i$, an orthonormal basis for the (real) $(d^2 - 1)$ dimensional vector space of traceless hermitian $d \times d$ matrices with the inner product $(X,Y) = \text{tr}(XY)$. For $\mathcal{U} \in SU(d)$, the components of the local operation in the correlation matrix basis, $U_{ij} = \text{tr}(\mathcal{U}\lambda_i \mathcal{U}^\dagger\lambda_j)$ is a special orthogonal matrix in $SO(d^2-1)$, representing $\mathcal{U}$ in the $(d^2 - 1)$ dimensional (adjoint) representation of $SU(d)$. More precisely,
the orthogonal matrices with components $U_{ij}$ form a subgroup of
$SO(d^2-1)$, being the homomorphic image of $SU(d)$ called the
adjoint group, which is the quotient of $SU(d)$ by its centre, the subgroup of matrices $\omega \mathbf{I}_d$ where $\omega$ is a $d$th root of unity. Thus, the property $U^T U=\mathbf{I}_d$ still holds and $\text{tr} S(\mathcal{C})$ is invariant under $SU(d)$. However, for $SL(d,\mathbb{C})$ the connection with a Lorentz group only works for $d = 2$.

We have seen for three pure qubits that only one
area enclosing path exists, but as the number of qubits increases,
there should be many more Kempe like quantities,
since there will be many more paths that enclose area. Our original
construction, on the other hand, produces many more invariants even
for three qubits, because it makes critical use of the `shrinking'
component of the transformation defined by the spin correlation
matrix.

One caveat with this approach as it stands is that we have only used the information contained in the two qubit density matrices. For states with large numbers of qubits one cannot obtain a full set of invariants since too much information about the overall state is lost when tracing out all but the two qubits in each link. This approach could be extended by considering contractions of the full correlation tensor. For example, an $N$ qubit system is described by a real tensor $S_{i_1 j_2\cdots z_N}$ where $i_1,j_2, \cdots ,z_N$ take the values $0,1,2,3$. It would be interesting to find an operational interpretation of contractions of these $N-$tensors.

%

\begin{acknowledgments}
We thank Johan {\AA}berg, Stephen Brierley, \v{C}aslav Brukner, Berge
Englert, Richard Jozsa, Noah Linden, Ognyan Oreshkov, Jiannis Pachos, Wonmin Son and Andreas Winter for helpful
comments and discussions. For financial support MSW acknowledges
EPSRC, QIP IRC www.qipirc.org (GR/S82176/01), the NRF and the MoE (Singapore) and an Erwin Schr\"{o}dinger JRF. ES and VV acknowledge the National Research Foundation and the Ministry of Education (Singapore). ME acknowledges support from the Swedish Research Council (VR).
\end{acknowledgments}

\bibliography{../../../references/masterbib}

\end{document}